\documentclass[a4paper]{aa}
\usepackage{psfig}
\usepackage{epsfig}
\usepackage{graphicx,float}
\usepackage{graphicx}

\def \xmm {XMM--Newton}
\def \sax {BeppoSAX}
\def \src {4U\thinspace1850$-$087}
\def \glob {NGC\thinspace6712}

\def \nh {N${\rm _H}$}

\def \hcm {\hbox {\ifmmode $ cm$^{-2}\else cm$^{-2}$\fi}}

\def \arcsec {\hbox{$^{\prime\prime}$}}
\def \chisq {$\chi ^{2}$}

\def\approxgt{\mathrel{\hbox{\rlap{\lower.55ex \hbox {$\sim$}}
        \kern-.3em \raise.4ex \hbox{$>$}}}}
\def\approxlt{\mathrel{\hbox{\rlap{\lower.55ex \hbox {$\sim$}}
        \kern-.3em \raise.4ex \hbox{$<$}}}}
\newcommand{\mc}{\multicolumn}

\begin{document}

\title{Low-energy absorption towards the ultra-compact
X--ray binary \src\ located in the globular cluster \glob}

\author{L. Sidoli\inst{1}
        \and N. La Palombara\inst{1}
    \and T. Oosterbroek\inst{2}
        \and A. N. Parmar\inst{3}
} \offprints{L. Sidoli, \email{sidoli@iasf.mi.cnr.it}}

\institute{Istituto di Astrofisica Spaziale e Fisica Cosmica ``G. Occhialini", IASF/INAF,
    via Bassini 15, I-20133 Milano, Italy
    \and
       Science Payload and Advanced Concepts Office, ESA, ESTEC,
       Postbus 299, NL-2200 AG, Noordwijk, The Netherlands
    \and
       Research and Scientific Support Department of ESA, ESTEC,
       Postbus 299, NL-2200 AG Noordwijk, The Netherlands
}
\date{Received 16 May 2005 / Accepted 2 July 2005}

\authorrunning{L. Sidoli et al.}

\titlerunning{Low-Energy absorption in \src}

\abstract{We report the results of two \xmm\ observations of the
ultra-compact low-mass X--ray binary \src\ located in the galactic
globular cluster \glob. A broad emission feature at 0.7~keV was
detected in an earlier ASCA observation and explained as the
result of an unusual Ne/O abundance ratio in the absorbing
material local to the source. We find no evidence for this feature
and derive Ne/O ratios in the range 0.14--0.21, consistent with
that of the interstellar medium. During the second observation,
when the source was $\sim$10\% more luminous, there is some
evidence for a slightly higher Ne/O ratio and additional
absorption. Changes in the Ne/O abundance ratio have been detected
from another ultra-compact binary, 4U\,1543$-$624. We propose that
these changes result from an X--ray induced wind which is
evaporated from an O and Ne rich degenerate donor. As the source
X--ray intensity increases so does the amount of evaporation and
hence the column densities and abundance ratio of Ne and O.
 }
\maketitle

\section{Introduction}
\label{sect:intro}

ASCA and BeppoSAX observations of ultra-compact (P$_{orb}$$<$1~hr)
low-mass X--ray binaries (LMXBs) have revealed two possible
spectral differences compared to the longer period systems. These
are (1) the presence of a discrete spectral feature near 0.7~keV
(attributed to Ne) in the ASCA spectra (Juett et al. \cite{j:01})
of X\thinspace1850$-$087, X\thinspace1543$-$624,
X\thinspace0614+01 and X\thinspace0918$-$549 (the orbital periods
of the last 2 sources are unknown, but their optical faintness is
consistent with an ultra-compact nature) and (2) their best-fit
parameter values when fitted with a disk-blackbody and Comptonized
continuum (Sidoli et al. \cite{s:01}). In the case of the
ultra-compact sources X\thinspace0512$-$401,
X\thinspace1820$-$303, X\thinspace1850$-$087 and
X\thinspace1832$-$330, fits to the BeppoSAX spectra give
significantly lower disk-blackbody temperatures than for other
LMXB. In addition, the Comptonization seed photon temperatures
appear consistent with those of the inner disk regions. Also 
the LMXBs located in the globular clusters NGC\,6652
(Parmar et al.~\cite{p:01}) and Terzan\,5 (Heinke et
al.~\cite{h:03}) show very similar spectral properties, suggesting
an ultra-compact nature for these binary systems (see
Verbunt~\cite{v:05} for a review).

The X--ray burst source X\thinspace1850$-$087 (Swank et al. 1976)
is an X--ray binary located in the galactic globular cluster
\glob, the least concentrated amongst those that host a luminous
X--ray source. A short period (20.6 minutes) UV modulation was
discovered with HST from the likely optical counterpart (Anderson
et al.~\cite{a:93}), implying a degenerate companion of
0.04$M_\odot$ (Homer et al.~\cite{h:96}). The source is located
$\sim$6\arcsec\ or 0.1$\pm{0.1}$ core radii from the cluster
center (Hertz \& Grindlay 1983).  Another UV--excess star was
discovered in the core of \glob\ with the ESO Very Large Telescope
(Ferraro et al. 2000), a few arcsec away from the LMXB. The
presence of these two interacting binaries inside the core of the
low-density cluster NGC\,6712 suggests that the interaction of the
cluster with the disk and bulge of our Galaxy during numerous
orbital passages plays a role in the formation of LMXBs in
globular clusters (Ferraro et al. 2000). Moreover, there is
evidence supporting a scenario where NGC\,6712 was much more
massive in the past and that it experienced a significant mass
evaporation produced by the tidal force due to interactions with
our Galaxy (see, e.g., Paltrinieri et al.~\cite{pal:01} and
references therein).

EXOSAT observations of \src\ revealed a complex spectrum. The
best-fit was obtained with a model consisting of a power-law with
a photon index, $\alpha$, of 0.4 with an exponential cut-off at
$\sim$1~keV, together with a blackbody with a temperature, $kT$,
of 2.4~keV and absorption, $N_{\rm H}$, of
$<$5$\times10^{21}$~\hcm\ (Parmar et al. 1989). A thermal
bremsstrahlung ($kT$ = 1.7~keV) absorbed by $5\times10^{21}$~\hcm\
is a good approximation to the ROSAT Position Sensitive
Proportional Counter spectrum (Verbunt et al. 1995). During a
\sax\ survey of the bright LMRXB located in galactic globular
clusters (Sidoli et al. \cite{s:01}) the 0.3--50~keV spectrum was
fit with a disk-blackbody and Comptonized continuum with $N_{\rm
H}$ = $3.9 \times 10^{21}$~\hcm, an inner disk temperature,
$kT_{\rm in}$, of 0.6 keV, an inner projected radius of $\sim$5~km
(for an assumed \glob\ distance of 6.8~kpc, Harris~\cite{ha:96}),
a temperature, $kT_{0}$, of the input ``seed'' photons, of 0.8 keV
(consistent with the inner disk-blackbody temperature), an
electron temperature, $kT_{\rm e}$, of 70~keV, and an optical
depth, $\tau$, of 1.7. The 0.1--100~keV luminosity was
1.9$\times10^{36}$~erg~s$^{-1}$.

Analysis of the ASCA Solid-state Imaging Spectrometer (SIS) data
(Juett et al. \cite{j:01}) from \src\ revealed the presence of a
spectral feature near 0.7~keV. A good fit to these data was found
with an absorbed $kT$ = 0.4~keV blackbody together with a
power-law with $\alpha$ = 2.1 when the relative abundances of O
and Ne were allowed to vary. Both components are absorbed by
$N_{\rm H}$ = 2.9$\times10^{21}$~\hcm, with a relative (to solar)
O/H abundance of 0.37$\pm{0.06}$ and Ne/H abundance of
1.9$\pm{0.3}$. The authors interpreted this excess absorption as
due to neutral Ne-rich material local to the binary. Preliminary
results from the 0.4--2~keV \src\ \xmm\ RGS spectra were reported
in Sidoli et al.~(\cite{s:04}), who found no evidence for an
anomalous Ne/O abundance ratio. Recently, analysis of {\em
Chandra} Low-Energy Transmission Grating Spectrometer (LETGS) data
confirmed this result (Juett \& Chakrabarty~\cite{j:05}),
measuring a Ne/O ratio of 0.22$\pm$0.05 consistent with that
expected from the interstellar medium (ISM) of 0.18 (Wilms et al.
2000).

Here we report the results of \xmm\ observations performed in
order to investigate the nature of the 0.7~keV feature found with
ASCA. We use the following updated values for the globular cluster
\glob\ parameters (Paltrinieri et al.~\cite{pal:01}): a distance
of 8$\pm{1}$~kpc (note that a previous estimate for the distance
was 6.8~kpc, Harris~\cite{ha:96}) and a reddening $E(B-V) = 0.33\pm{0.05}$.
Adopting the relation $A_{\rm V} = 3.1\, E(B-V)$, and $A_{\rm V} =
N_{\rm H} \times 5.59 \times 10^{-22}$~\hcm\ 
(Predehl \&
Schmitt~\cite{ps:95}) the optical reddening translates into an ISM
column density of (1.8$\pm{0.2}$)$\times$10$^{21}$~\hcm\ to \glob.

\begin{table*}
\caption{\xmm\ on-axis observation log of \src. Two observations
were performed. The MOS1, MOS2 and pn cameras all used the medium
thickness filter.} \label{tab:log}
\begin{tabular}[c]{llcccll}
\hline\noalign{\smallskip}
Obs.  &  Start time     & End time    & Inst. &\mc{1}{c}{Net Exp.} & Mode  \\
      & (dy~mon~yr~hr:mn)     & (dy~mon~yr~hr:mn) &       &\mc{1}{c}{(ks)}     &       \\
\noalign{\smallskip\hrule\smallskip}
1     &  27 Sep 2003 09:05 &  27 Sep 2003 12:22 & MOS1    & 11.7   &  Timing Fast Uncompressed  \\
1     &  27 Sep 2003 09:05 &  27 Sep 2003 12:22 & MOS2    & 11.5   &  Imaging Prime Partial Window \\
1     &  27 Sep 2003 09:10 &  27 Sep 2003 12:23 & pn      &  8.1   &  Imaging Small Window  \\
1     &  27 Sep 2003 09:04 &  27 Sep 2003 12:28 & RGS1    & 12.1   &  Spectroscopy  \\
1     &  27 Sep 2003 09:04 &  27 Sep 2003 12:28 & RGS2    & 12.1   &  Spectroscopy  \\
\noalign{\smallskip\hrule\smallskip}
2     &  09 Oct 2003 08:15 &  09 Oct 2003 10:35 & MOS1    &  8.3   &  Timing Fast Uncompressed  \\
2     &  09 Oct 2003 08:15 &  09 Oct 2003 10:39 & MOS2    &  8.4   &  Imaging Prime Partial Window \\
2     &  09 Oct 2003 08:20 &  09 Oct 2003 10:41 & pn      &  5.9   &  Imaging Small Window  \\
2     &  09 Oct 2003 08:14 &  09 Oct 2003 10:43 & RGS1    &  8.8   &  Spectroscopy  \\
2     &  09 Oct 2003 08:14 &  09 Oct 2003 10:43 & RGS2    &  8.8   &  Spectroscopy  \\
\noalign{\smallskip\hrule\smallskip}
\end{tabular}
\end{table*}

\section{Observations}
\label{sect:obs}

The XMM-Newton Observatory (Jansen et al. \cite{ja:01}) includes
three 1500~cm$^2$ X--ray telescopes each with an European Photon
Imaging Camera (EPIC) at the focus. Two of the EPIC imaging
spectrometers use MOS CCDs (Turner et al.~\cite{t:01}) and one
uses a pn CCD (Str\"uder et al. \cite{st:01}). Behind two of the
telescopes there are Reflection Grating Spectrometers (RGS,
0.35--2~keV; den Herder et al. \cite{dh:01}). \xmm\ observed \src\
twice, due to visibility problems, in 2003 September and October,
about 12~days apart (see Table~\ref{tab:log} for the observation
details).

Data were reprocessed using version 6.1 of the Science Analysis
Software (SAS). Known hot, or flickering, pixels and electronic
noise were rejected using the SAS. The latest response matrices
were used (updated to 2004-12-03, which should improve the
agreement between the MOS and pn below 1 keV,
Saxton~\cite{sx:04}), while the ancillary response files were
generated using the SAS task {\em arfgen}. Spectra were selected
from single events only (pattern 0) for the MOS1 timing mode (only
pattern 0 has been calibrated in this instrument mode) while for
MOS2 patterns from 0 to 12 and for the pn patterns from 0 to 4
were selected. Source counts were extracted from circular regions
of 40\arcsec\ radius centered on \src\ for the pn and the MOS2.
With the SAS task {\em epatplot} we verified that pn Small Window
data are not significantly affected by pile-up, whereas in both
MOS spectra pile-up was evident. Thus, we minimized the effects of
pile-up by extracting MOS2 events in an annulus outside of a
10\arcsec\ radius core of the \src\ point spread function, and
MOS1 events from a wide column outside the central 15\arcsec. A
comparison between MOS1 and MOS2 spectra revealed that after this
selection the source spectral shapes observed by the two
instruments were similar. We use the pn for the determination of
the source flux. Background counts were obtained from similar
regions offset from the source position. The backgrounds do not
show any evidence for flaring activity, so the entire nominal
exposure times were considered. For both observations, the RGS
spectra were analyzed as produced by the pipeline processing
performed by XMM-Newton Survey Science Centre.

In order to ensure applicability of the \chisq\ statistic, the
extracted spectra were rebinned such that at least 20 counts per
bin were present and such that the energy resolution was not
over-sampled by more than a factor 3. Note that no systematic
uncertainties were added to the spectra. All spectral
uncertainties and upper-limits are given at 90\% confidence for
one interesting parameter.

\section{Results}

\hspace{1.5cm}
\begin{figure*}
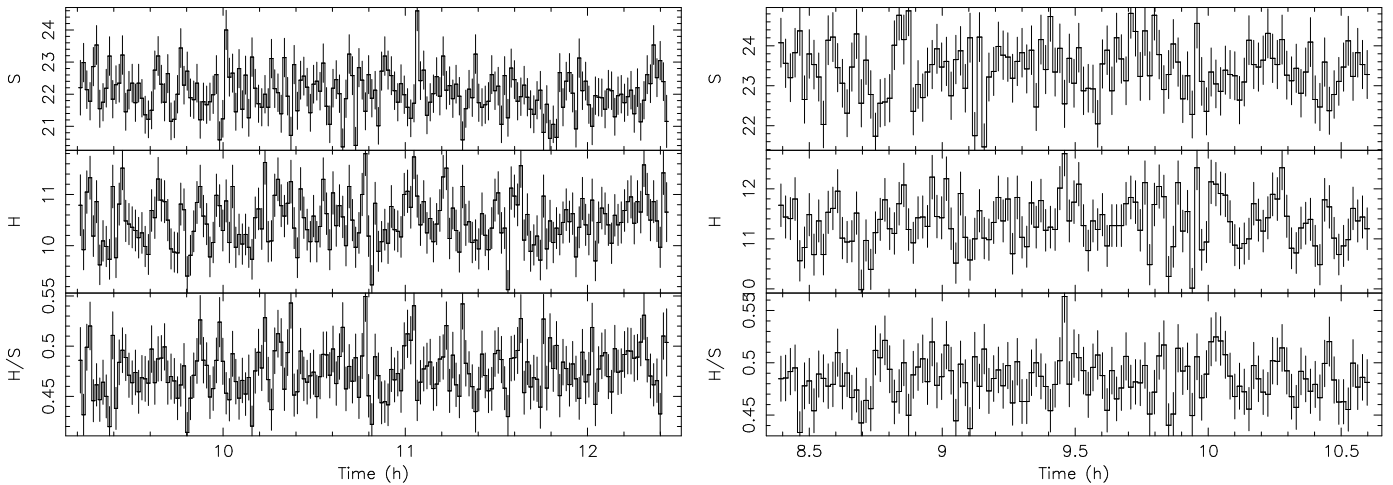

\hbox{\hspace{0.cm}
\includegraphics[height=9.0cm,angle=-90]{lc1.ps}
\hspace{.1cm}
\includegraphics[height=9.0cm,angle=-90]{lc2.ps}}
\caption[]{EPIC pn lightcurves in two energy bands (S, 0.3--2 keV
and H, 2--10 keV) and hardness ratios, H/S, during the first
(left) and second observation (right). Time is in hours of 2003
September 27 (for the first observation) and of the 2003 October 9
(for the second observation). The binning is 64~s.} \label{fig:lc}
\end{figure*}

\subsection{Lightcurves}

\begin{figure}[h!]
\includegraphics[height=9.0cm,angle=-90]{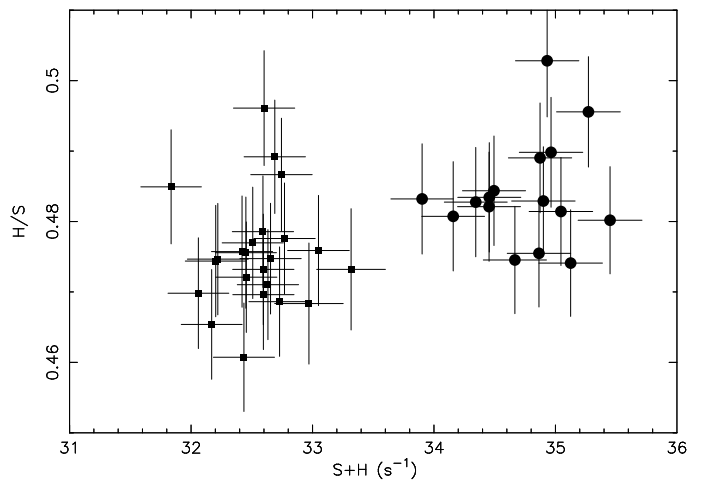}
\caption[]{Hardness ratio (H = 2--10~keV; S = 0.3--2~keV counts)
versus total intensity (S+H, 0.3--10~keV) for the two pn observations
(squares mark the first observation, circles the second). The
binning is 512~s. } \label{fig:hard}
\end{figure}

Lightcurves for the two observations in soft (0.3--2 keV) and hard
(2--10 keV) energy ranges were extracted in order to search for
variability and hardness ratio variations (Fig.~\ref{fig:hard}).
The source was slightly harder and more intense during the second
observation. Thus, the two observations were analyzed separately.
Since within each individual observation the source does not show
evidence for intensity or hardness variations (see
Fig.~\ref{fig:lc}), we can safely consider two separate spectra
extracted from each of the two observations without making any
further selections. The statistical quality of the data, combined
with the length of the observations, does not allow for a
meaningful search for periods around the optical period.

\subsection{Spectra}

We performed separate spectral analysis for the two \xmm\
observations. We first studied the pn, MOS1 and MOS2 spectra in
the energy range 0.3--12~keV. We noticed that the pn spectrum, in
both the observations, showed a significant departure from the
MOS1 and MOS2 shapes, especially around 0.6~keV, where a large
excess is present only in the residuals of the pn spectrum
(Fig.~\ref{fig:comparison}). A significant departure of the pn
spectrum in this energy range is present also compared with the
RGS1 and RGS2. Note that this excess is not at the same energy of
the feature present in the ASCA SIS spectra, which was interpreted
as due to additional absorption by neutral Ne. Smaller differences
between the pn and MOS cameras are also present at other energies,
below 0.4~keV, and up to about 1.7~keV. Uncertainties in the
calibration of the pn Small Window mode below 2~keV are reported
in Kirsch et al.~(\cite{k:04}), although the use of the latest
updated response matrices should reduce these differences
(Saxton~\cite{sx:04}). Examination of the pn background spectra
does not reveal any features at these energies, so we are
confident that they are not due to improper background
subtraction. Moreover, both observations show similar shapes for
the structured residuals. Thus, we restricted the pn energy range
to 1.7--12~keV, where there are no significant differences between
the MOS1, MOS2 and pn spectra.

In summary, we used the following energy ranges: 0.4--2~keV for
RGS1 and RGS2, 0.3--8~keV for the MOS1 (because of the low
statistics at high energy), 0.3--10~keV for the MOS2, and
1.7--12~keV for the pn. The RGS spectra in both observations do
not show evidence for edges or emission features. We investigated
the 0.3--12~keV \src\ spectra by simultaneously fitting the RGS1,
RGS2, MOS1, MOS2 and pn spectra of each of the two observations
individually. Factors were included in the spectral fitting to
allow for normalization uncertainties between the instruments. In
the spectral fitting {\sc xspec} version 11.2 was used, and the
interstellar abundances of Wilms et al.~(\cite{w:00}) were used in
the photoelectric absorption models.

\vskip 0cm
\begin{small}
\begin{figure}[h!]
\centering
\includegraphics[width=5.0cm,angle=-90]{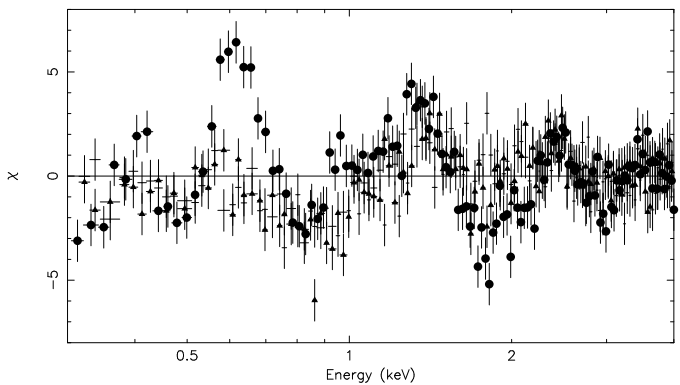}
\caption{Residuals (in units of standard deviations) when fitting
the 2003 September MOS and pn spectra with a single absorbed
power-law model. There is a different structure of the residuals
in the pn (filled circles), compared with MOS1 (crosses) and MOS2
(triangles). A similar structured excess is present in the 2003
October spectrum. } \label{fig:comparison}
\end{figure}
\end{small}

LMXB X--ray spectra are generally fit with two component models, a
black-body or a multicolor disk-blackbody component to account for
the low-energy emission (originating from the accretion disk or
the neutron star surface), and a high-energy component which is
usually modeled with power-law, cut-off power-law, or Comptonized
components, to account for the high-energy emission thought to be
produced in a corona. In order to check if a soft component was
required by the data we tried first with the simplest model
consisting of an absorbed power-law (photoelectric absorption
model {\sc phabs} in {\sc xspec}). The fits resulted in positive
residuals between 0.3--0.7~keV and reduced $\chi^2$ = 1.69 for
1098 degrees of freedom (d.o.f.), and $\chi^2$ = 1.94 for 969
d.o.f. for the first and the second observations, respectively.
Adding a blackbody improved the fits (reduced $\chi^2$ = 1.51 for
1096 d.o.f, and $\chi^2$ = 1.78 for 967 d.o.f.), but structured
residuals at low-energies remain.
Using a disk-blackbody component instead of the
blackbody resulted in better fits, with reduced $\chi^2$=1.49 for
1096 d.o.f., and $\chi^2$=1.76 for 967 d.o.f. for the first and
second observations, respectively.

The absorption resulting from these fits was always higher than
that derived from the optical reddening to \glob. As well as the
interstellar absorption in the direction of the globular cluster,
modeled with {\sc phabs} with N${\rm _H}$ fixed at
1.8$\times$10$^{21}$~\hcm, we added another multiplicative
component to investigate whether an ionized absorber could be
present ({\sc absori} model in {\sc xspec}). We fixed the Fe
abundance of the absorber to the \glob\ value ([Fe/H]=$-0.80\pm{0.2}$),
and linked the photon index of the ionizing continuum to that of
the power-law component. The fit resulted in an un-ionized cold
absorbing medium, with a best-fit ionization $\xi = L/nR^2$ in the
range 0.5--2 (where $L$ is the ionizing luminosity, $n$ the
density of the absorbing medium, and $R$ is the distance of the
obscuring material to the source). Thus, we do not consider
further the presence of any ionized absorber.

We next tried partial covering ({\sc pcfabs} model in {\sc
xspec}), absorbing both the disk-blackbody and power-law continuum
components. We included a {\sc phabs} component with $N{\rm _H}$
fixed at 1.8$\times$10$^{21}$~\hcm\ to account for the
interstellar absorption. The fit resulted in additional absorption
in the range 6--8$\times$10$^{21}$~\hcm\ for the two observations,
with a covering factor $\sim$95\% for reduced $\chi^2$ = 1.37 for
1095 d.o.f., and $\chi^2$ = 1.40 for 966 d.o.f. for the first and
the second observations, respectively. The disk-blackbody
temperatures were 0.6 and 0.3~keV (with the innermost radii of the
accretion disk ${r \rm _{in}({\cos}i)^{0.5}}$ = 3.9$\pm{0.5}$~km
and 17$\pm{6}$~km) while $\alpha$ = 2.22$\pm{0.05}$ and
2.35$\pm{0.02}$ (with power-law normalizations of 0.060$^{+0.001}
_{-0.008}$ and 0.090$\pm{0.003}$
photons~keV$^{-1}$~cm$^{-2}$~s$^{-1}$ at 1~keV, for the first and
the second observations, respectively).

\begin{table*}[!ht]
\caption[]{Best-fit continuum parameters when the RGS1, RGS2, pn,
MOS1 and MOS2 \src\ spectra were fit simultaneously. {\em
dbb}=disk-blackbody, {\em pow}=power-law, {\em ctt}=Comptonization
model {\sc comptt} in {\sc xspec}. 
$\alpha$ is the
power-law photon index, $kT_{\rm in}$ the inner disk temperature,
$r{\rm _{in}}(\cos i)^{0.5}$ is the inner disk radius for a
distance of 8~kpc, $i$ is the disk inclination angle and $R{\rm
_{bb}}$ is the blackbody radius. For the {\sc comptt} model,
$kT{\rm _0}$ is the temperature of the ``seed'' photons, $kT{\rm
_e}$ is the electron temperature (fixed at 70~keV) and ${\rm
\tau_p}$ is the plasma optical depth.}
\begin{center}
\begin{tabular}[c]{lclllllllr}
\hline\noalign{\smallskip} Model  &\mc{1}{c}{$N{\rm _H}$}
&\mc{1}{c}{$\alpha$} &\mc{1}{c}{$kT{\rm _{in}}$} &\mc{1}{c}{$r{\rm
_{in}} (\cos i)^{0.5}$} &\mc{1}{c}{$kT{\rm _{bb}}$}
&\mc{1}{c}{$R{\rm _{bb}}$} &\mc{1}{c}{$kT{\rm _0}$}
&\mc{1}{c}{${\rm \tau_p}$}
& $\chi^2$/dof \\
  &($10^{21}$~cm$^{-2}$)&   &\mc{1}{c}{(keV)} &\mc{1}{c}{(km)}
&\mc{1}{c}{(keV)} &\mc{1}{c}{(km)}
&\mc{1}{c}{(keV)} & &  \\
\noalign{\smallskip\hrule\smallskip}
\mc{3}{l}{Observation 1}    & & &  & & & &   \\
dbb+pow        & $5.7 ^{+0.3}_{-0.2}$ & $2.03^{+0.06}_{-0.06}$   & $0.69^{+0.02}_{-0.02}$  & $2.9^{+0.2}_{-0.2}$  &\dots & \dots & \dots & \dots  & 1.34/1092 \\
bb+pow            & $6.1^{+0.2}_{-0.2} $          & $2.18^{+0.04}_{-0.04}$    & \dots                          & \dots              & $0.48^{+0.02}_{-0.02}$ & $5.1 ^{+0.6}_{-0.6} $& \dots& \dots&  1.37/1092 \\
dbb+ctt    & 4.1$^{+0.1}_{-0.1}$   & \dots & $0.73  ^{+0.03}_{-0.02}$& $2.5  ^{+0.3}_{-0.2}$& \dots& \dots& $0.23^{+0.01} _{-0.03}$&$1.2^{+0.1}_{-0.1}$ &  1.30/1091 \\
\noalign{\smallskip\hrule\smallskip}
\mc{3}{l}{Observation 2}  & & & & & & &   \\
dbb+pow        & $5.7^{+0.2}_{-0.3}$ & $2.16  ^{+0.06}_{-0.06}$   & $0.75^{+0.06} _{-0.04}$ & $1.8  ^{+0.4}_{-0.4}$  & \dots & \dots &  \dots & \dots   & 1.43/963 \\
bb+pow            & $5.9^{+0.2}_{-0.2}$        & $2.24 ^{+0.04}_{-0.04}$   & \dots                     & \dots              & $0.55^{+0.07}_{-0.04}$ & 2.7$ ^{+1.0}_{-0.7}$ & \dots &  \dots &  1.44/963 \\
dbb+ctt     & $3.6^{+0.2}_{-0.5}$ & \dots  & $0.84^{+0.10} _{-0.05}$ & $1.5 ^{+0.4}_{-0.4}$ & \dots& \dots& $0.22^{+0.03} _{-0.02}$ &  $1.0^{+0.1} _{-0.1}$&  1.38/962 \\
\noalign{\smallskip\hrule\smallskip}
\end{tabular}
\label{tab:spec}
\end{center}
\end{table*}

\begin{figure*}[ht!]
\hbox{\hspace{0.9cm}
\includegraphics[height=5.7cm,angle=0]{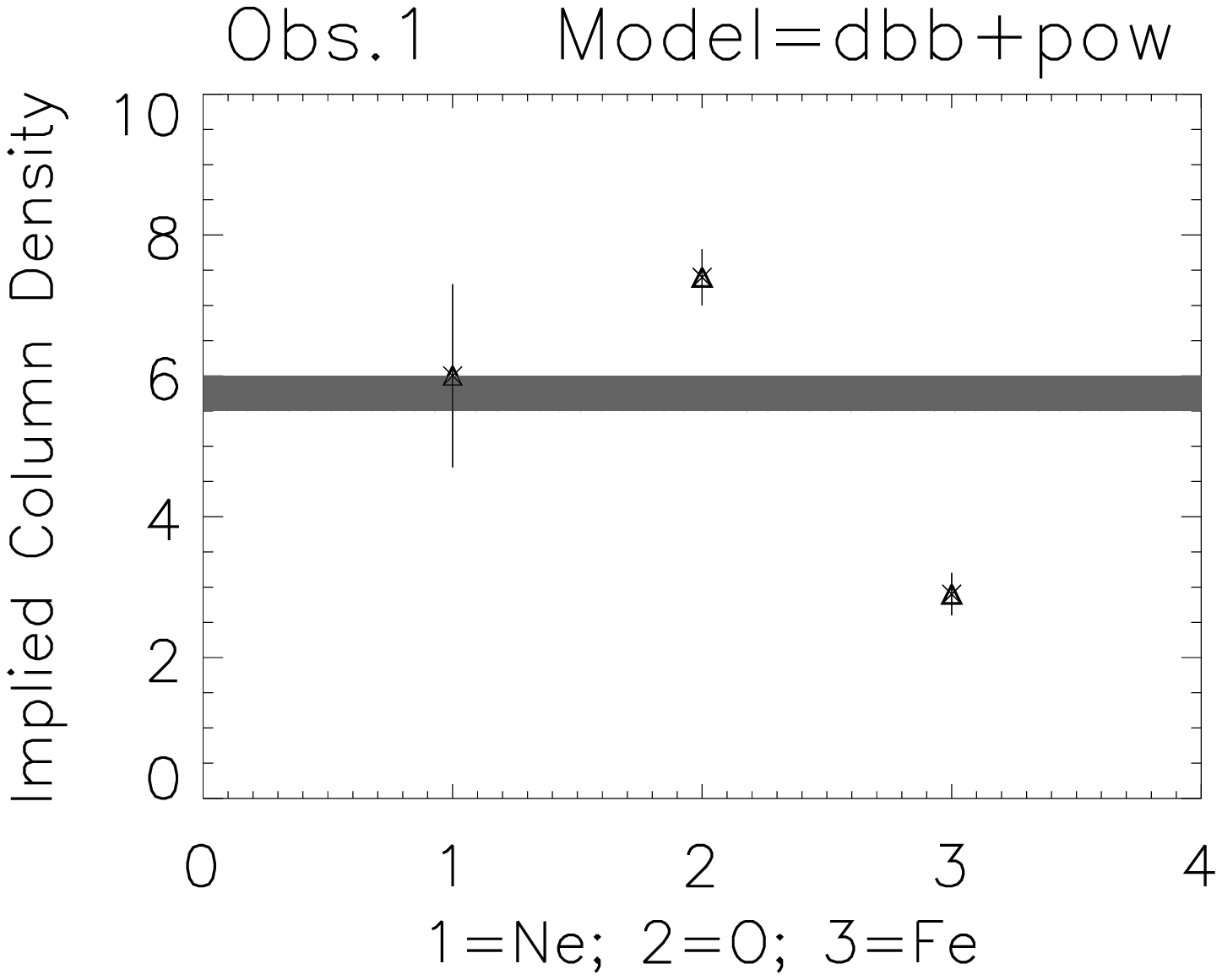}
\hspace{1.2cm}
\includegraphics[height=5.7cm,angle=0]{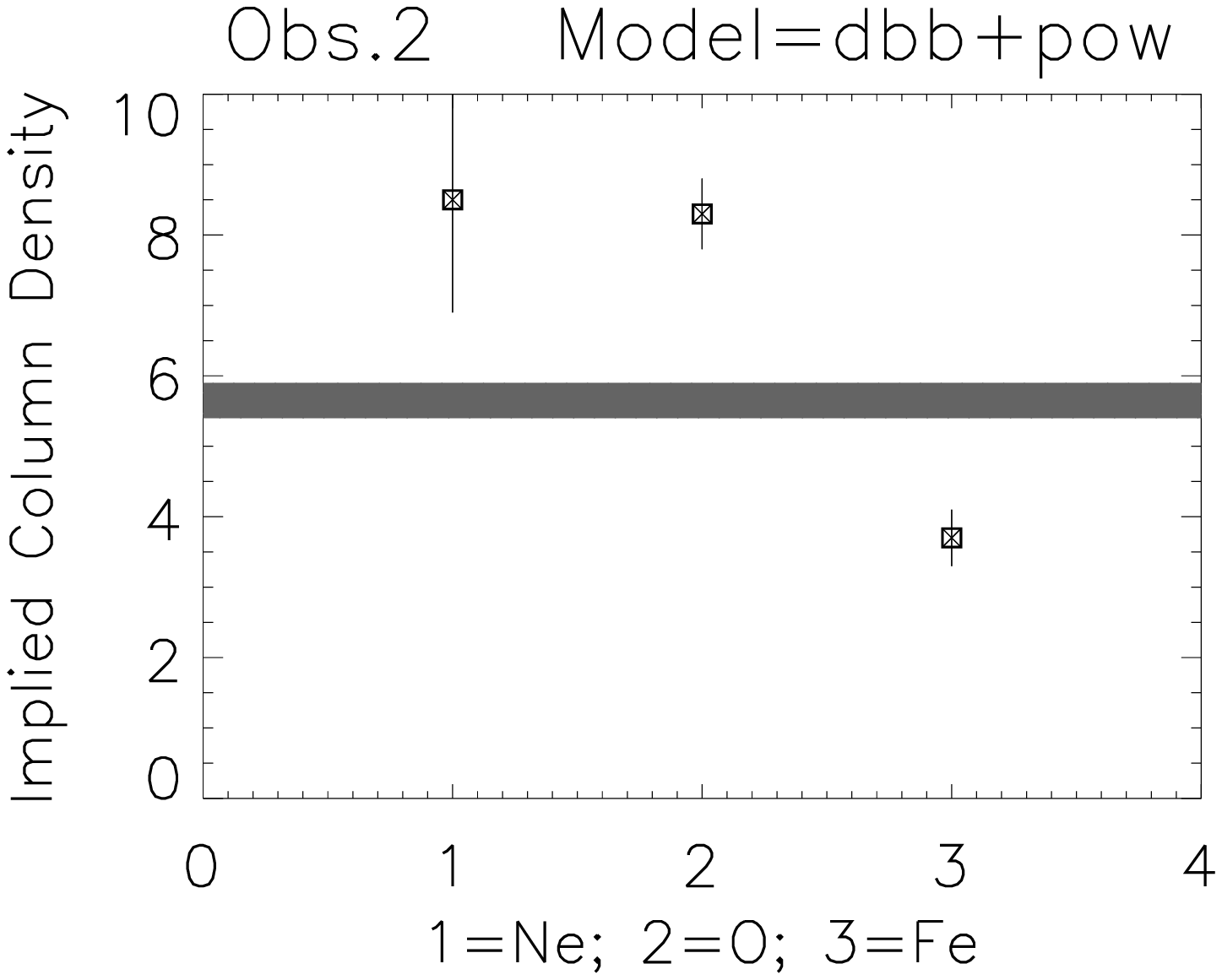}}
\vbox{\vspace{0.15cm}}

\hbox{\hspace{0.9cm}
\includegraphics[height=5.7cm,angle=0]{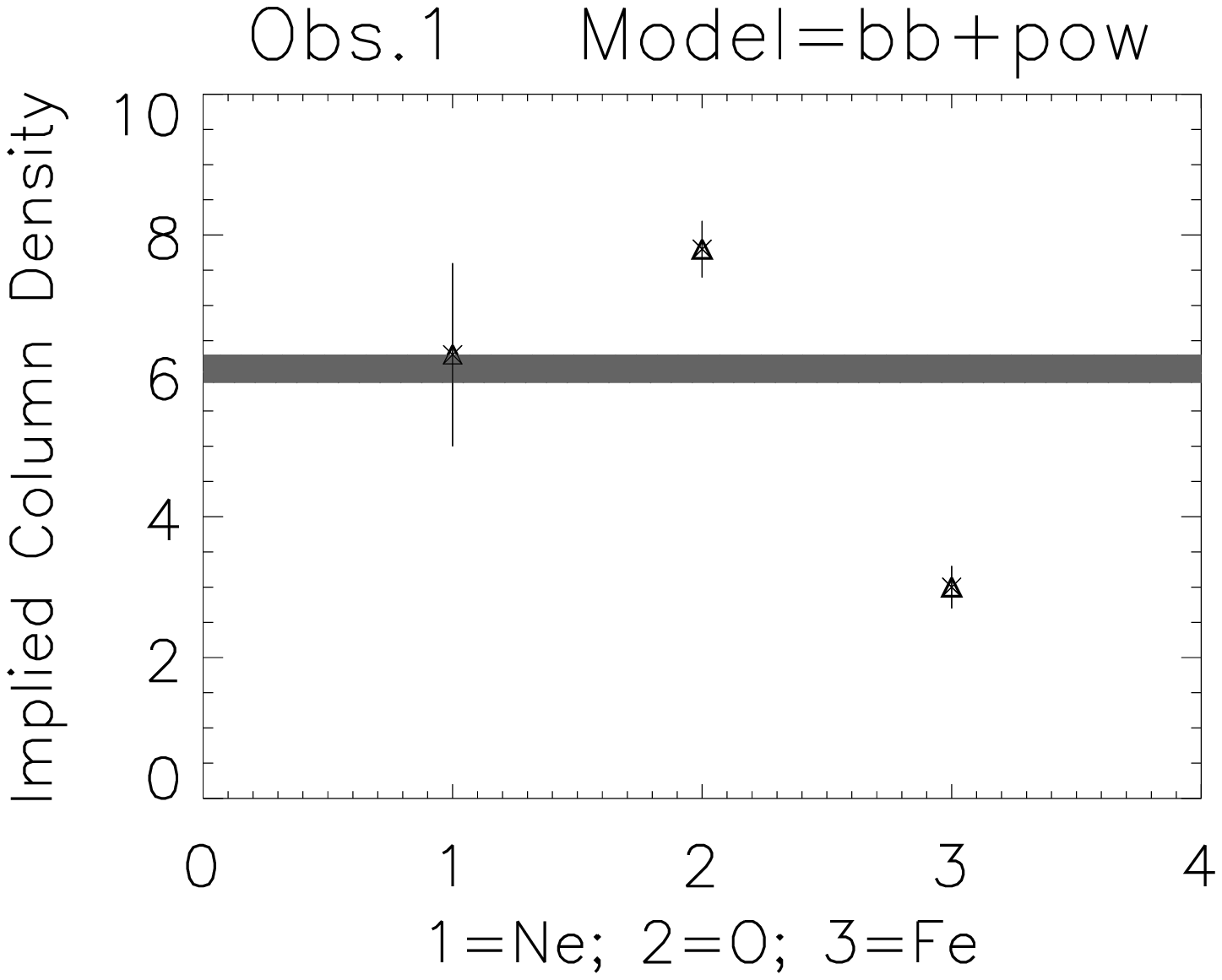}
\hspace{1.2cm}
\includegraphics[height=5.7cm,angle=0]{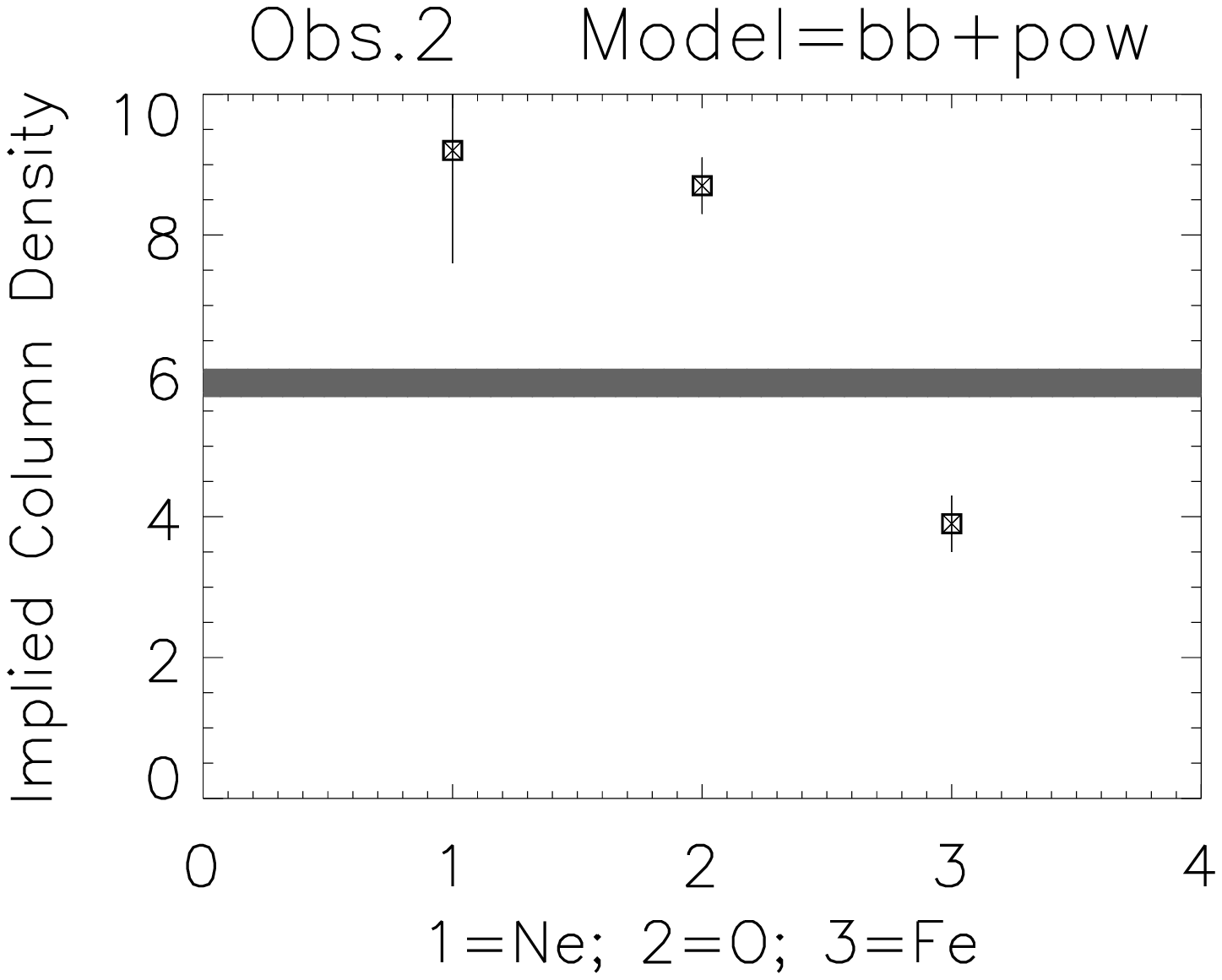}}
\vbox{\vspace{0.15cm}}

\hbox{\hspace{0.9cm}
\includegraphics[height=5.7cm,angle=0]{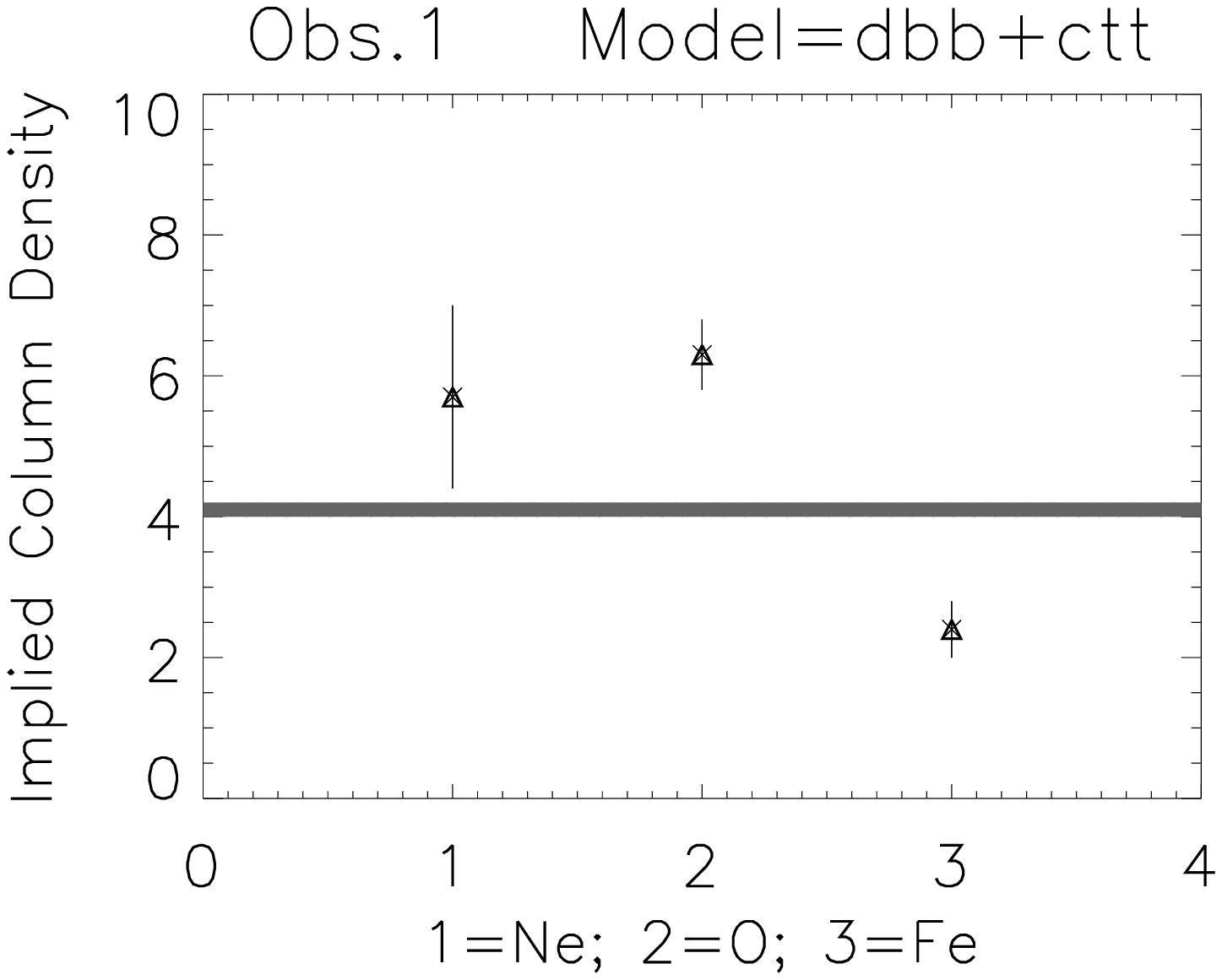}
\hspace{1.2cm}
\includegraphics[height=5.7cm,angle=0]{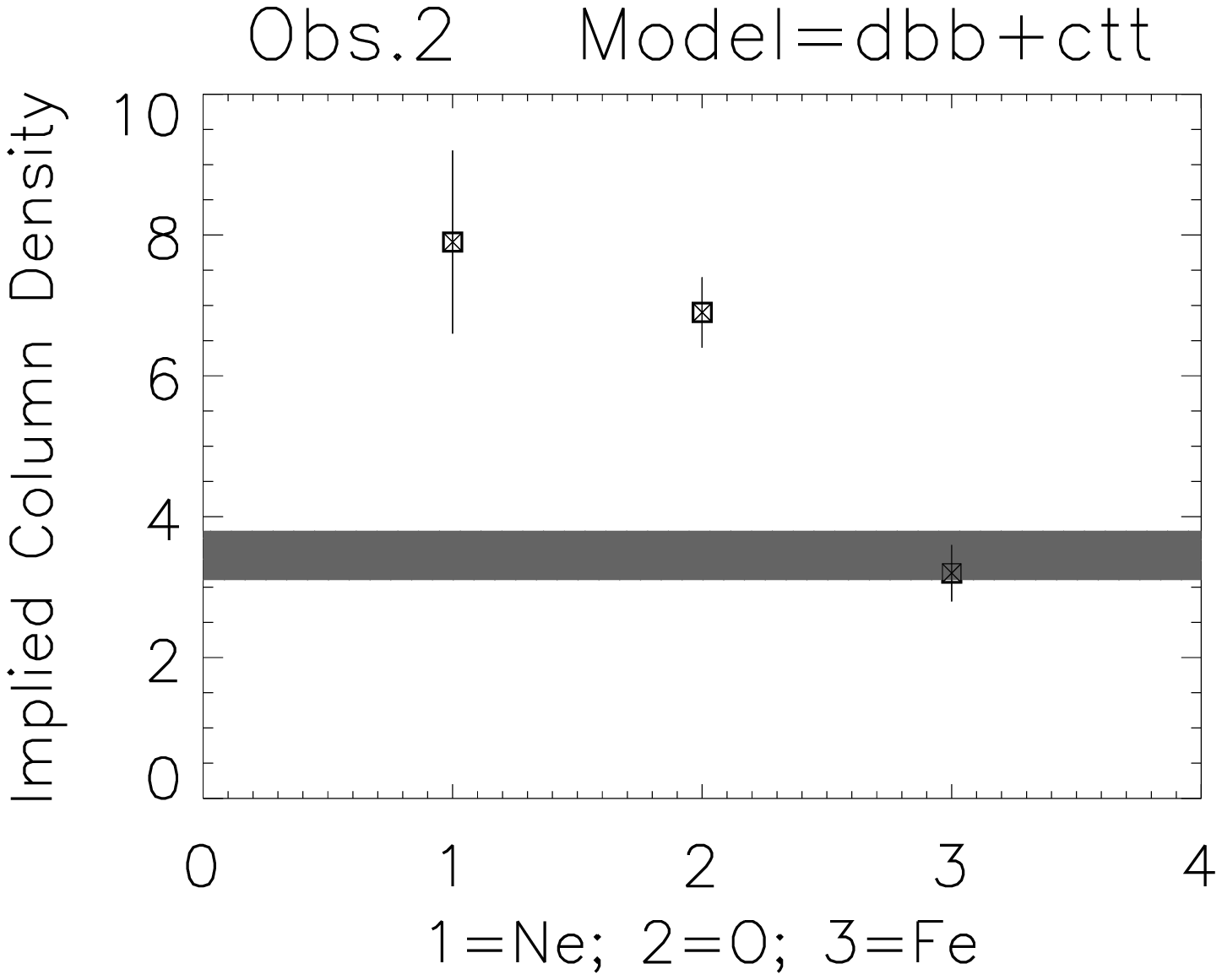}}

\caption[]{The hydrogen column densities (in units of
10$^{21}$~\hcm) implied by the optical depths of the Ne, O and Fe
edges, from the first (left) and second observations (right),
depending on the different models assumed for the continua
(Table~\ref{tab:edge}). Shadowed regions mark the 90\% confidence
range for $N_{\rm H}$ resulting from the continuum fits ({\sc
vphabs} model, see Table~\ref{tab:spec}). } \label{fig:nh}
\end{figure*}

\begin{table}[!ht]
\caption[]{Photoelectric absorption results towards \src\ (see
Table~\ref{tab:spec} for the continuum emission parameters). The
edge energies were fixed at 0.54, 0.71 and 0.87~keV for O, Fe and
Ne, respectively. $\tau_{\rm edge}$ is the absorption depth.
$N_{\rm Z}$ is the element column density (in units of
10$^{17}$~\hcm) calculated using the Henke et al.~(\cite{he:93})
cross sections. $N_{\rm H}$ is the hydrogen column density implied
by $N_{\rm Z}$, in units of 10$^{21}$~\hcm, assuming the ISM
abundances of Wilms et al.~(\cite{w:00}).}
\begin{center}
\begin{tabular}[c]{llllll}
\hline\noalign{\smallskip}
Model  & \mc{1}{c}{Edge} &\mc{1}{c}{$\tau_{\rm edge}$} &\mc{1}{c}{$N_{\rm Z}$}      &\mc{1}{c}{$N_{\rm H}$} &\mc{1}{c}{Ne/O} \\
\noalign{\smallskip\hrule\smallskip}
\multicolumn{3}{l}{Observation 1}  \\
\cline{1-2}
dbb+    & Ne K & $0.19^{+0.04} _{-0.05}$ &$5.2^{+1.1}_{-1.4}$  &$6.0^{+1.3}_{-1.6}$ & \\
pow        & O K  & $2.07^{+0.12}_{-0.12}$  & $36.4^{+2.1}_{-2.1}$  & $7.4^{+0.4}_{-0.4}$  & $0.14^{+0.03}_{-0.04}$ \\
           & Fe L & $0.54^{+0.06}_{-0.06}$ & $0.77^{+0.09}_{-0.09}$ & $2.9^{+0.3}_{-0.3}$  &  \\
\hline
bb+   & Ne K & $0.20 ^{+0.04} _{-0.04}$ & $5.5^{+1.1}_{-1.1}$  & $6.3^{+1.3}_{-1.3}$ &   \\
pow        & O K  & $2.17 ^{+0.12} _{-0.12}$ & $38.2^{+2.1}_{-2.1}$  & $7.8^{+0.4}_{-0.4}$ & $0.14^{+0.03}_{-0.03}$  \\
           & Fe L & $0.57 ^{+0.06} _{-0.06}$ & $0.81^{+0.09}_{-0.09}$  & $3.0^{+0.3}_{-0.3}$  &   \\
\hline
dbb+    & Ne K & $0.18 ^{+0.04} _{-0.05}$ & $5.0^{+1.1}_{-1.1}$  & $5.7^{+1.3}_{-1.6}$& \\
+ctt    & O K  & $1.76 ^{+0.14} _{-0.13}$ & $31.0^{+2.5}_{-2.5}$ & $6.3^{+0.5}_{-0.5}$ & $0.16^{+0.04}_{-0.05}$    \\
        & Fe L & $0.46 ^{+0.07} _{-0.06}$ & $0.66^{+0.10}_{-0.10}$& $2.4^{+0.4}_{-0.3}$ &  \\
\noalign{\smallskip\hrule\smallskip}
\multicolumn{3}{l}{Observation 2}  \\
\cline{1-2}
dbb+    & Ne K & $0.27 ^{+0.05}_{-0.05}$  & $7.4^{+1.4}_{-1.4}$  & $8.5^{+1.6}_{-1.6}$ &    \\
pow        & O K  & $2.32 ^{+0.13}_{-0.13}$ &$40.8^{+2.3}_{-2.3}$& $8.3^{+0.5}_{-0.5}$&$0.18^{+0.04}_{-0.04}$ \\
           & Fe L & $0.70 ^{+0.07}_{-0.07}$ &$1.0^{+0.1}_{-0.1}$& $3.7^{+0.4}_{-0.4}$ &   \\
\hline
bb+     & Ne K & $0.29 ^{+0.05}_{-0.05}$ & $8.0^{+1.4}_{-1.4}$ &$9.2^{+1.6}_{-1.6}$  &   \\
pow     & O K  & $2.41 ^{+0.12}_{-0.12}$ & $42.4^{+2.1}_{-2.1}$& $8.7^{+0.4}_{-0.4}$ & $0.19^{+0.03}_{-0.03}$  \\
        & Fe L & $0.73 ^{+0.07}_{-0.07}$ & $1.0^{+0.1}_{-0.1}$  & $3.9^{+0.4}_{-0.4}$ &   \\
\hline
dbb+    & Ne K & $0.25 ^{+0.04} _{-0.08}$ & $6.9^{+1.1}_{-2.2}$  & $7.9^{+1.3}_{-2.5}$  &   \\
ctt     & O K & $1.90 ^{+0.14} _{-0.22}$ & $33.5^{+2.5}_{-3.9}$  & $6.9^{+0.5}_{-0.8}$ & $0.21^{+0.04}_{-0.07}$ \\
        & Fe L & $0.60 ^{+0.07} _{-0.11}$ & $0.86^{+0.10}_{-0.16}$  & $3.2^{+0.4}_{-0.4}$ &  \\
\noalign{\smallskip\hrule\smallskip}
\end{tabular}
\end{center}
\label{tab:edge}
\end{table}

In order to investigate the Ne/O abundance ratio, we used a
variable abundance absorption model ({\sc vphabs} in {\sc xspec}),
with the elemental abundances set to the ISM values of Wilms et
al.~(\cite{w:00}) except for those of O, Ne and Fe 
which were fixed to zero. Their absorption effect has been replaced
with three edges (O-K, Fe-L, Ne-K edges) with energies fixed at
0.54, 0.71 and 0.87 keV, and edge depths allowed to vary.
In this way we could also account for a local iron abundance likely
different from the cosmic value. 
We fit the
spectra with 3 different two-components models for the continuum:
(1) a disk-blackbody ({\sc diskbb} in {\sc xspec}, Mitsuda et
al.~\cite{m:84}) and a power-law, (2) a blackbody and a power-law,
and (3) a disk-blackbody and a high-energy Comptonized component
({\sc comptt} in {\sc xspec}, Titarchuk \cite{ti:94}). This latter
model was considered since it has been successfully fit to almost
{\em all} the broad-band spectra of the galactic globular cluster
LMXBs (Sidoli et al. \cite{s:01}). Since the \sax\ best-fit
electron temperature, $kT_{\rm e}$, of $\sim$70~keV is well above
the XMM-Newton upper energy threshold, it was fixed to 70~keV for
the fits performed here.

\begin{figure*}
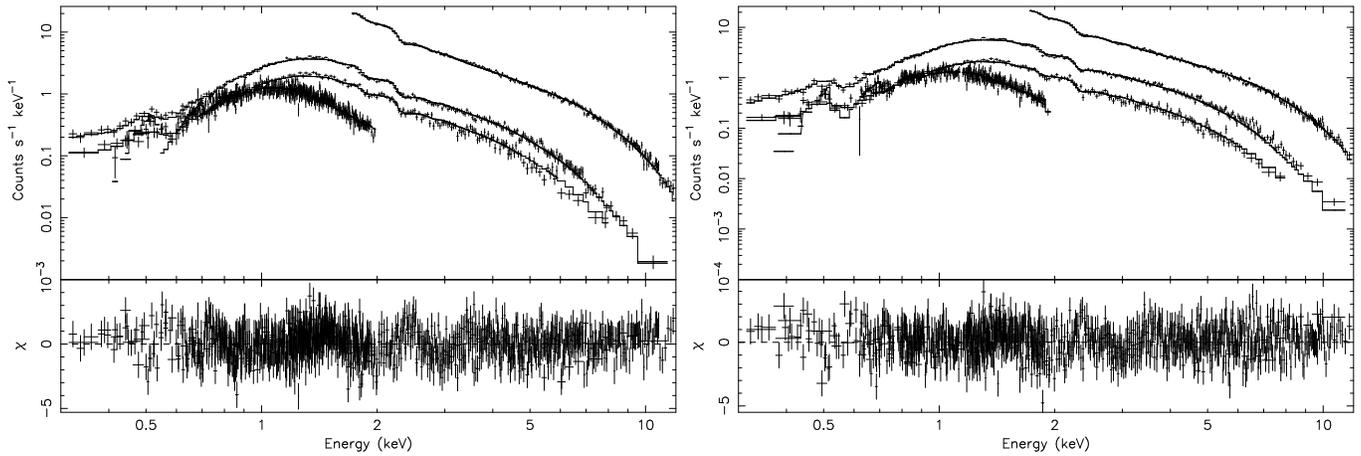

\hbox{\hspace{.1cm}
\includegraphics[height=8.9cm,angle=-90]{rebinned_201.ps}
\includegraphics[height=8.9cm,angle=-90]{rebinned_501.ps}}
\caption[]{The 0.3--12~keV  \src\ count spectra (together with the
residuals in units of standard deviation) from the two
observations (left panel shows the first observation, the right
panel the second). The continua have been modeled with a
disk-blackbody and a power-law (see Tables~\ref{tab:spec} and
\ref{tab:edge} for the continuum and line parameters,
respectively).}
 \label{fig:bestfit}
\end{figure*}

All 3 continuum models fit well and the best-fit continuum
parameters are given in Table~\ref{tab:spec}. All 3 models give
similar 0.5--10~keV source luminosities of $1.4 \times
10^{36}$~erg~s$^{-1}$ and $1.6 \times 10^{36}$~erg~s$^{-1}$ for a
distance of 8~kpc for the first and second observations,
respectively. Table~\ref{tab:edge} gives the measured Ne/O
abundance ratios. These are all similar to the ISM value of 0.18
of Wilms et al. (2001). The equivalent column densities due to O,
Ne and Fe were then estimated by modeling the absorption as
occurring from 3 edges with energies fixed at 0.54, 0.71 and
0.87~keV together with a narrow absorption line at an energy of
0.53~keV (to account for O~I ISM absorption). The resulting edge
depths, columns and equivalent hydrogen columns are listed in
Table~\ref{tab:edge} and shown in Fig.~\ref{fig:nh}.
In Fig.~\ref{fig:bestfit} we show the best fit spectra 
during the two observations.

\begin{figure*}
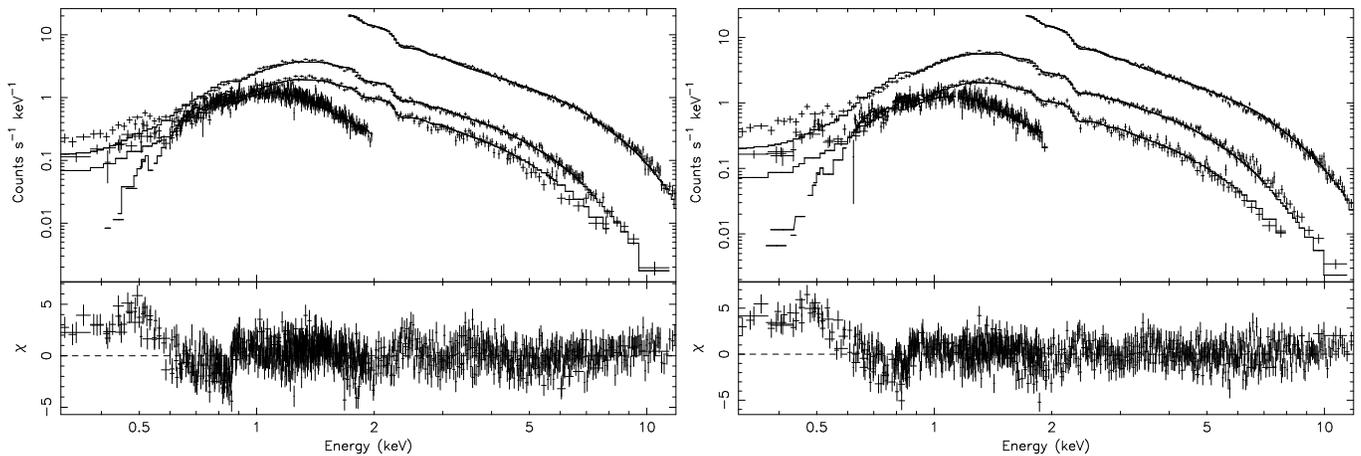

\hbox{\hspace{.1cm}
\includegraphics[height=8.9cm,angle=-90]{asca_rebinned_201_5.ps}
\includegraphics[height=8.9cm,angle=-90]{asca_rebinned_501_5.ps}}
\caption[]{The 0.3--12~keV \src\ count spectra (together with the
residuals in units of standard deviations) from the two
observations (left panel shows the first observation, the right
panel the second), fit with the ASCA model with the Ne and O
abundances fixed at the values reported in Juett et
al.~(\cite{j:01}).}
 \label{fig:asca}
\end{figure*}

As a final test, we tried fitting the spectra with the ASCA model,
with the Ne and O abundances fixed at the ASCA best-fit values 
(Juett et al.~\cite{j:01}), letting all
the other parameters vary in the usual way. We obtained
unacceptable fits with reduced $\chi^2$ = 2.0 for 1096 d.o.f., and
$\chi^2$ = 2.2 for 967 d.o.f. with structured residuals evident
below 1~keV (see Fig.~\ref{fig:asca}).

\section{Discussion}
\label{sect:discussion}

We report the results of two \xmm\ observations of \src\ performed
about 12 days apart. The spectra require a soft emission
component, which is slightly better described by a multi-color
disk-blackbody than a blackbody. At higher energies, a power-law
provides a good fit the spectra. The photon index is similar to
that measured during the ASCA observation (Juett et
al.~\cite{j:01}), while it is significantly softer than during the
later $Chandra$ observation (Juett \& Chakrabarty \cite{j:05}).
This may be due to the fact that in fitting the $Chandra$ LETGS
spectrum the low-energy absorption was fixed to the ISM value
towards the globular cluster. We note however, that the source
luminosities during the \xmm\ and $Chandra$ observations were
similar, whilst during the ASCA observation the source was almost
a factor 2 brighter with a 0.5--10 keV luminosity of $2.5 \times
10^{36}$~erg~s$^{-1}$.

The total low-energy absorption resulting from the fits is similar
in both the \xmm\ observations and is 4--6.3$\times$10$^{21}$
~\hcm, depending on the model adopted for the continuum. There is
evidence for extra-absorption in the line of sight, since the
best-fit total \nh\ is always significantly higher than the
optically derived value in the direction of the host globular
cluster of ($1.8\pm 0.2)\times$10$^{21}$ ~\hcm. Thus the intrinsic
absorption ranges from 2 to 4.5$\times$10$^{21}$ ~\hcm, depending
on the continuum model assumed. In the Comptonization model a
lower column density is required because of the turnover at low
energies present in the model. The presence of neutral
extra-absorption local to the source is also confirmed by a good
fit  when using the partial covering fraction absorption model,
which indicate that the central source is absorbed by a neutral
medium with a covering factor of $\sim$95\% and an intrinsic
hydrogen column density in the range 6--8$\times$10$^{21}$~\hcm\
for the two observations.

We adopted a variable absorption model together with three edges
(O-K, Ne-K and Fe-L) in order to measure the column density of the
Ne, O and Fe in the line of sight, since the ASCA spectrum suggests
an excess absorption of neutral Ne-rich material local to the
source (Juett et al.~\cite{j:01}). Other ultra-compact X--ray
binaries ($P_{\rm orb}$$<$1~hr) display over-abundances of neutral
Ne from the absorption effects in ASCA spectra. Moreover, during
XMM-Newton and $Chandra$ observations an anomalously high Ne/O
abundance ratio has been observed in a a number of other
ultra-compact binaries, indicative of neutral Ne overabundance
(e.g., 4U\,0614+091, Paerels et al.~\cite{pa:01}; 4U\,1543--624
and 2S\,0918--549, Juett \& Chakrabarty~\cite{j:03}). Thus, it has
been proposed that the donor stars in some ultra-compact binary systems
are Ne-rich white dwarfs (e.g., Yungelson et al.~\cite{y:02};
Bildsten~\cite{b:02}).

The high resolution RGS \src\ spectra presented here do not show
any prominent emission features or other absorption edges, besides
those of O-K, Ne-K and Fe-L. Our study of the absorbing Ne and O
toward \src\ reveals an Ne/O abundance ratio (see
Table~\ref{tab:edge}) which is consistent with the ISM value of
0.18 (Wilms et al.~\cite{w:00}). This result is in agreement with
the $Chandra$ observation performed in 2002 (Juett \& Chakrabarty,
\cite{j:05}) but contrary to the earlier ASCA measurement (Juett
et al. \cite{j:01}).

If the measured elemental column densities  $N_{\rm Z}$ (see
Table~\ref{tab:edge}) are converted to equivalent H column
densities using the Wilms et al.~(\cite{w:00}) ISM abundances, we
can compare the $N{\rm _H}$ resulting from the overall shape of
the \xmm\ spectra (dashed regions in Fig.~\ref{fig:nh} include the
90\% uncertainties on the $N{\rm _H}$ resulting from the fit). For
each continuum model, for both observations, there seems to be a
discrepancy between the total \nh\ and the equivalent $N {\rm _H}$
calculated from the elemental column densities $N_{\rm Z}$. This
could indicate an over-abundance of Ne and O and a sub-solar
abundance of Fe. This under-abundance may be explained by the low
Fe abundance of the host globular cluster. Alternatively, it is
possible that the uncertainties on the derived hydrogen column
densities could be underestimated. Indeed, Paerels et al. (2000)
point out that the photoelectric cross-sections could have a 30\%
uncertainty which is large enough to account for this discrepancy.

There is no evidence for any temporal variability of the Ne, O and
Fe column densities between the two \xmm\ observations (within
90\% uncertainty), although we note that the column densities are
systematically higher in the second observation. Also the Ne/O
ratio, although always compatible within uncertainties with the
standard ISM value, is on average higher during the second
observation. The \src\ spectrum during the two \xmm\ observations
differs in the total 0.5--10~keV luminosity (see
Table~\ref{tab:spec}) with the second observation being $\sim$10\%
more luminous than the first. This suggests a possible correlation
between the Ne/O abundance ratio and the X--ray source luminosity,
and a possible explanation for the different Ne/O ratios observed
with \xmm, $Chandra$, and ASCA. Note that during the ASCA
observation, where evidence for a strong Ne overabundance was
reported, the source luminosity was almost twice that during the
\xmm\ and $Chandra$ observations.

Juett \& Chakrabarty~(\cite{j:03}) proposed that ionization could
play a role in the variable abundance ratios. 
We suggest another
possible mechanism which could help in the understanding of the
local extra-absorption and its metal abundance. Maccarone et
al.~(\cite{m:04}) studied the irradiation-induced stellar winds in
X--ray binaries in order to explain different X--ray spectra from
LMXBs located in globular clusters with different metallicities.
An evaporative wind can be produced even in LMXBs with degenerate
companions, where part of the radiation produced from the central
source illuminates the donor star (e.g., Ruderman et
al.~\cite{r:89}). This wind could contribute to the observed
column density toward the X--ray source, leading to an intrinsic
column density of $\sim$6$\times10^{21}$~\hcm\ (see eq.~(7) in
Maccarone et al.~\cite{m:04}). During the \xmm\ observations we
found a comparable amount of extra-absorption towards \src. We
suggest that a wind evaporated from the degenerate companion could
be responsible for the intrinsic absorption observed. Using the
\src\ observed parameters, and eq.~(7) of Maccarone et
al.~(\cite{m:04}), we derive a wind velocity of
$\sim$5$\times10^{7}$~cm~s$^{-1}$, which may be confined to the
binary (the escape velocity is $\sim$$10^{8}$~cm~s$^{-1}$). A
higher source luminosity would translate into a larger
contribution by the wind from the degenerate donor, which is
likely to be rich in Ne and O. We suggest that this mechanism
could contribute to the different abundance ratio observed from
\src\ with \xmm, $Chandra$, and ASCA. This could possibly help in
explaining why during higher source luminosity intervals, higher
Ne/O abundance ratios are observed. We note that another
ultra-compact X--ray binary, 4U\,1543--624, displays a variable
Ne/O abundance ratio. Juett \& Chakrabarty (2003) measured an Ne/O
abundance ratio of 1.5$\pm{0.3}$ with $Chandra$, and
0.54$\pm{0.03}$ with \xmm. The higher Ne/O abundance ratio was
observed when the source was more luminous, as appears to be the
case with \src.

Theoretical models for the formation of white dwarfs
predict a Ne/O abundance ratio in the range 0.2--0.4 
(e.g., Deloye \& Bildsten~\cite{db:02},
Segretain et al.~\cite{se:94}, Gutierrez et al.~\cite{gu:96}),
which is low compared with the Ne/O ratios observed with ASCA in \src\ (or 
with $Chandra$ in 4U\,1543--624).
On the other hand,  Yungelson et al.~\cite{y:02}, studying
the formation of Ne-enriched donors in ultracompact X--ray binaries,
point out that the abundance of neon in the nucleus of the dwarf 
may be well underestimated by a factor of 3 (Isern et al.~\cite{i:91}), 
which makes the theoretically predicted Ne/O ratios agree better
with the higher observed values.

\begin{acknowledgements}
Based on observations obtained with XMM-Newton, an ESA science
mission with instruments and contributions directly funded by ESA
member states and the USA (NASA).
\end{acknowledgements}


\begin{thebibliography}{}

\bibitem[1993]{a:93}
Anderson, S. F., Margon, B., Deutsch, E. W., \& Downes, R. A.
1993, AJ, 106, 1049

\bibitem[2002]{b:02}
Bildsten, L. 2002, ApJ, 577, L27


\bibitem[2002]{db:02}
Deloye, C. J., Bildsten, L., 2002, ApJ, 580, 1077


\bibitem[2001]{dh:01}
Den Herder, J. W., Brinkman, A. C., Kahn, S. M., et al. 2001, A\&A, 365, L7

\bibitem[2000]{f:00}
Ferraro, F. R., Paltrinieri, B., Paresce F., \& De Marchi, G.
2000, ApJ, 542, L29

\bibitem[1987]{f:87}
Frank, J., King, A. R., \& Lasota, J.-P. 1987, A\&A, 178, 137


\bibitem[1996]{gu:96}
Gutierrez, J., Garcia-Berro, E., Iben, I., et al., 1996, ApJ, 459, 701



\bibitem[1996]{ha:96}
Harris, W. E. 1996, AJ, 112, 1487

\bibitem[2003]{h:03}
Heinke, C., Edmonds, P., Grindlay, J., et al. 2003, ApJ, 590, 809

\bibitem[1993]{he:93}
Henke, B. L., Gullikson, E. M., \& Davis, J. C. 1993, Atomic Data
\& Nucl. Data Tables, 54, 181

\bibitem[1983]{hg:83}
Hertz, P., \& Grindlay, J. E. 1983, ApJ, 275, 105

\bibitem[1996]{h:96}
Homer, L., Charles, P. A., Naylor, T., et al. 1996, MNRAS, 282,
L37

\bibitem[1991]{i:91}
Isern, J.,  Hernanz, M., Mochkovitch, R., et al., 1991, A\&A, 241, L29

\bibitem[2001]{ja:01}
Jansen, F., Lumb, D., Altieri, B., et al. 2001, A\&A, 365, L1

\bibitem[2001]{j:01}
Juett, A. M., Psaltis, D., \& Chakrabarty, D. 2001, ApJ, 560, L59

\bibitem[2003]{j:03}
Juett, A. M., \& Chakrabarty, D. 2003, ApJ, 599, 498

\bibitem[2005]{j:05}
Juett, A. M., \& Chakrabarty, D. 2005, ApJ in press
(astro-ph/0501472)

\bibitem[2004]{k:04}
Kirsch, M. G. F., Altieri, B., Chen, B., et al. 2004, {\em
http://xmm.vilspa.esa.es/docs/documents/CAL-TN-0055-1-0.ps.gz},
(astro-ph/0407257)

\bibitem[2004]{m:04}
Maccarone, T. J., Kundu, A., \& Zepf, S. E. 2004, ApJ, 606, 430

\bibitem[1984]{m:84}
Mitsuda, K., Inoue, H., Koyama, K., et al. 1984, PASJ, 36, 741

\bibitem[2001]{pa:01}
Paerels, F., Brinkman, A. C., van der Meer, R. L. J., et al. 2001,
ApJ, 546, 338

\bibitem[1989]{p:89}
Parmar, A. N., Stella, L., \& Giommi, P. 1989, A\&A, 222, 96

\bibitem[2001]{p:01}
Parmar, A. N., Oosterbroek, T., Sidoli, L., et al. 2001, A\&A,
380, 490

\bibitem[1995]{ps:95}
Predehl, P., \& Schmitt, J. H. M. M. 1995, A\&A, 293, 889

\bibitem[2001]{pal:01}
Paltrinieri, B., Ferraro, F. R., Paresce, F., et al. 2001, AJ,
121, 3114

\bibitem[1989]{r:89}
Ruderman, M., Shaham, J., Tavani, M., et al. 1989, ApJ, 343, 292

\bibitem[2004]{sx:04}
Saxton, R. D., 2004, \xmm\ CCF Release Note 183, {\em
http://xmm.vilspa.esa.es/docs/documents/CAL-SRN-0183-1-0.ps.gz}


\bibitem[1994]{se:94}
Segretain, L., Chabrier, G., Hernanz, M., et al., 1994, ApJ, 434, 641


\bibitem[2001]{s:01}
Sidoli, L., Parmar, A. N., Oosterbroek, T., et al. 2001, A\&A,
368, 451

\bibitem[2004]{s:04}
Sidoli, L., Parmar, A. N., \& Oosterbroek, T. 2004, Proc. of the V
INTEGRAL Workshop, ESA SP-552, p.~389

\bibitem[2001]{st:01}
Str\"uder, L., Briel, U., Dennerl, K., et al. 2001, A\&A, 365, L18

\bibitem[1976]{s:76}
Swank, J. H., Becker, R. H., Pravdo, S. H., et al. 1976, IAU Circ.
3010

\bibitem[1994]{ti:94}
Titarchuk L. 1994, ApJ, 434 570

\bibitem[2001]{t:01}
Turner, M. J. L., Abbey, A., Arnaud, M., et al. 2001, A\&A, 365, L27

\bibitem[2005]{v:05}
Verbunt, F. 2005, Proc. of ``Interacting binaries", July 4-10
Cefalu, eds. Antonelli, L., et al., to be published with
AIP.(astro-ph/0412524)

\bibitem[1995]{v:95}
Verbunt, F., Bunk, W., Hasinger, G., \& Johnston, H. M. 1995,
A\&A, 298, 21

\bibitem[2000]{w:00}
Wilms, J., Allen, A., \& McCray, R. 2000, ApJ, 542, 914

\bibitem[2002]{y:02}
Yungelson, L. R., Nelemans, G., \& van den Heuvel, E. P. J. 2002,
A\&A, 388, 546

\end{thebibliography}
\end{document}